\documentclass[11pt]{article}
\usepackage{graphicx}
\input{epsf}
\begin{document}
\pagenumbering{arabic}
\begin{center}
\title{ ESTIMATION OF SIGNAL AND NOISE PARAMETERS FROM CMB POLARIZATION
OBSERVATIONS}
\author{\hspace{0.5cm}{\bf P. Naselsky$^{1,2}$, D.Novikov$^{3,4}$, 
I.Novikov$^{1,4,5,6}$ and J.Silk$^{3}$}.\\
\small $^1$ Theoretical Astrophysics Center, Copenhagen , Denmark. \\ 
\small $^2$ Rostov State University,  Rostov-Don, Russia.\\
\small $^3$ Astronomy Department, University of Oxford, UK. \\
\small $^4$ Astro-Space Center of P.N.Lebedev Physical Institute, 
\small Moscow, Russia. \\
\small $^5$ Astronomical University Observatory,  Copenhagen , Denmark.\\
\small $^6$ NORDITA, Copenhagen , Denmark.}

\date{}
\maketitle
\end{center}
\begin{abstract}

We propose a technique for determination of the spectral parameters
of the cosmological signal and pixel noise using observational
data on CMB polarization without any additional assumptions.
We introduce the notion of so called crossing points in the
observational maps and derive the theoretical dependence of the
total number of crossing points at different level of the polarization.
Finally, we use the statistics of the signal in the
vicinities of the singular points, where the polarization of the
pure CMB vanishes to correct the final result.

\vspace{0.3cm}

{\it Subject headings:} cosmic microwave background, 
cosmology, statistics, observations.
\end{abstract}

\section{Introduction}

Recent observational data by  BOOMERANG and MAXIMA-1 (De Bernardis et al.
2000, Hanany et al. 2000) opened 
a new epoch 
in the investigation of the CMB power spectrum at large multipole 
numbers $l$. 
The angular power spectrum of the CMB  (in particular the positions and 
amplitudes of the first, second and subsequent Doppler peaks) provides a 
unique 
chance to construct the most likely cosmological model. This model 
includes information about the Hubble constant, baryonic 
and CDM densities, cosmological constant $\Omega_{\Lambda}$, 
ionisation history and so on. 
In the coming years the measurements of the angular anisotropy of CMB
by satellite missions will provide CMB maps with high
resolution and sensitivity. 

In addition to the anisotropy of the
intensity it is possible, though more difficult, to measure polarization
of the radiation. Polarization is a secondary effect induced by the
scattering of anisotropic radiation on electrons in the cosmic plasma.
Importance of the polarization measurements of the 
relict radiation  was pointed out by Rees (1968) and this problem 
has since been discussed in 
many papers.
Polarization contains an addition to the anisotropy information about the
nature of primordial cosmological perturbations and different types
of foregrounds. The polarization field is a combination of two randomly 
distributed Stokes parameters ($Q$ and $U$) while the anisotropy is just 
a scalar. In particular, this field is quite sensitive to the
presence of tensor perturbations and a deviation from zero of the so called
pseudo scalar or 'magnetic' part of polarization would be an indicator
of gravitational waves or vector perturbations.  
 
Analogously to the anisotropy of the CMB, one of the major problems 
in the future analysis of polarization maps is the 
separation of the noise from the original cosmological signal.
Most of the denoising techniques require significant assumptions
about expected signal and noise to be made before the data analyzed. 
One example of
such technique is Wiener filtering (Tegmark and Efstathiou 1996, 
Bouchet and Gispert 1999). We would like to focus our attention
on the following problem: is it possible to find the spectral
parameters of at least some kind of noise using the observational 
data without any additional assumptions. In this case we could
use the real spectral parameters instead of the assumed one 
for the subsequent filtering.
We use geometrical and statistical properties of the CMB polarization
field for the following purposes:\\
1. To find the parameters of the signal and pixel noise;\\
2. To detect noise in the regions of the map where
polarization vanishes.

For solving the first problem we suggest investigation of 
so called up-crossing and down-crossing points of the modulus of polarization
as well as $Q$ and $U$ components separately at different levels in the 
pixelized map. For the pure CMB signal
these points are situated along the isopolarization lines and their 
number is proportional
to the total length of these lines. In this case the length of such 
lines is known analytically. 
In the 
presence of pixel noise these 'lines' become wider and are completely
destroyed (look like spots) in the vicinities of zero points where the 
signal is smaller than the noise. The analytical formula (derived by us)
for the number of these points in the case of the presence of Gaussian CMB signal
and pixel noise gives us a unique possibility to find the spectral parameters
of signal and noise with high accuracy.  

The second part of our investigation is the natural generalization of the first one.
We show that it is useful to
study the singular points in polarization where polarization
is vanishing. Such singular points of polarization have the 
common property that in the
vicinity of each point the polarization field is formed mainly by a small
scale noise (pixel noise and (or) point sources). Noise could 
manifest itself due to influence on the weak CMB signal in the vicinities 
of non-polarized points.

\section{General properties of the polarization field}

Here we will describe some general properties of CMB polarization and 
present the necessary formalism. We make a simplifying assumption that the
relevant angular scales are sufficiently small, so that  
the corresponding part on the sky is almost flat.
In this approximation the polarization field on the sky can be considered 
as a two dimensional field on the $(x,y)$-plane. 
Since Thompson scattering does not produce circular polarization, the
resulting field can be completely described in terms of two Stokes parameters
$Q$ and $U$. Without loss of generality we can consider a cosmological model
with scalar perturbations only. In this case parameters $Q$ and $U$ are determined
by a single scalar field $\phi$. These parameters can be written in the 
following form:

\begin{equation}
\begin{array}{l}
Q=\frac{\partial^2\phi}{\partial x^2}-\frac{\partial^2\phi}{\partial y^2},\\
U=2\frac{\partial^2\phi}{\partial x\partial y},
\end{array}
\end{equation}
where $\phi(x,y)$ is supposed to be in the form of a random Gaussian field.
For further investigations we have to introduce the spectral parameters
as follows:

\begin{equation}
\begin{array}{l}
\sigma_0^2=\langle Q^2\rangle=\langle U^2\rangle, \\
\sigma_1^2=\langle Q_x^2\rangle=\langle U_x^2\rangle=
\langle Q_y^2\rangle=\langle U_y^2\rangle. \\
\end{array}
\end{equation}
Using these terms one can write the joint probability distribution functions (PDF)
for the fields $Q$, $U$, $P$ ($P=\sqrt{Q^2+U^2}$) and their first derivatives
in x-direction (see for details (Naselsky and D.Novikov, 1998)). 
Since PDF for Q and U are identical, 
we restrict ourself to theoretical
investigation of $Q$ and $P$ fields only:

\begin{equation}
\begin{array}{l}
F_Q(Q,Q_x)dQdQ_x=\frac{1}{2\pi\sigma_0\sigma_1}e^{-\frac{Q^2}{2\sigma_0^2}}
e^{-\frac{Q_x^2}{2\sigma_1^2}}dQdQ_x,\\
F_P(P,P_x)dPdP_x=\frac{P}{\sqrt{2\pi}\sigma_0^2\sigma_1}e^{-\frac{P^2}{2\sigma_0^2}}
e^{-\frac{P_x^2}{2\sigma_1^2}}dPdP_x.
\end{array}
\end{equation} 

Let us consider the behavior of a smooth continuous two-dimensional 
random field f(x,y) in the direction x with y kept fixed. We define the crossing point as the
point where this field crosses some threshold $\nu$. In the small vicinity of this point
$df=f_xdx$ and $f=\nu+f_x\Delta x$. Therefore, the probability $\Delta$ 
to find such a point between $x$ and $x+\Delta x$ is:

\begin{equation}
\begin{array}{l}
\Delta^+=\Delta x\int\limits^{+\infty}_{0}F(\nu,f_x)f_xdf_x \hspace{1cm}    
up-crossing,\\
\Delta^-=-\Delta x\int\limits_{-\infty}^{0}F(\nu,f_x)f_xdf_x \hspace{1cm} 
down-crossing,\\
\Delta=\Delta^++\Delta^-=\Delta x\int\limits_{-\infty}^{+\infty}F(\nu,f_x)|f_x|df_x.
\end{array}
\end{equation} 
Using equations (3,4) we can easily find density of crossing points
for Stokes parameters $\Delta_Q$ and for the modulus of polarization $\Delta_P$:

\begin{equation}
\begin{array}{l}
\Delta_q=\frac{1}{2\pi}\frac{\Delta x}{r_c}e^{-\frac{q^2}{2}},\\
\Delta_p=\frac{p}{\sqrt{2\pi}}\frac{\Delta x}{r_c}e^{-\frac{p^2}{2}}.
\end{array}
\end{equation} 
Here, we use dimensionless values $q=Q/\sigma_o$, $p=P/\sigma_o$ and
$r_c=\sigma_o/\sigma_1$ is the correlation radius.

In the two dimensional map these points are along the isolines 
of  $q$ or $p$ 
respectively and the density
of such points is proportional to the total length of these lines in
the map (fig. 1). This is one of the so called Minkowski functionals (see
Schmalzing and Gorski, 1997).

\section{CMB polarization and noise in the pixelized map.}

\begin{center}
{\it 3.1 Definitions}
\end{center}

The real observational datasets have a pixelized form. This means that we
should consider the field f(x,y) which is defined at the points $x_i$, $y_j$.
Without loss of generality one can use rectangular map $N\times N$ pixels
with distance $h$ between them: $x_{i+1}-x_{i}=h$. In this case $\Delta x$ in
the formulae (5) should be replaced by h. The field f crosses some threshold
$\nu$ between two neighbor pixels (i,j) and (i+1,j) if
$f_{i,j}<\nu$ and $f_{i+1,j}>\nu$ (up-crossing) or $f_{i,j}>\nu$ and 
$f_{i+1,j}<\nu$ (down-crossing). The position of the crossing point $x_{cr}$
can be defined by the linear interpolation of the field between two neighboring
pixels: 

\begin{equation}
\begin{array}{l}
x_{cr}=x_i+h\frac{\nu -f_{i,j}}{f_{i+1,j}-f_{i,j}},\\
y_{cr}=y_j.
\end{array}
\end{equation}
Analogously we find the positions of crossing points along the y-direction.
Finally, we construct the map with  the set of crossing points that are placed on the 
grid lines (fig.1). This set of points obviously form lines of the same level for the field
(if this field is smooth enough, namely if $h<<r_c$ or (the same) $h\sigma_1<<\sigma_0$).

\parbox{5.0in}
   
        {\includegraphics [scale=0.5,width=4.09in,totalheight=2.602in]{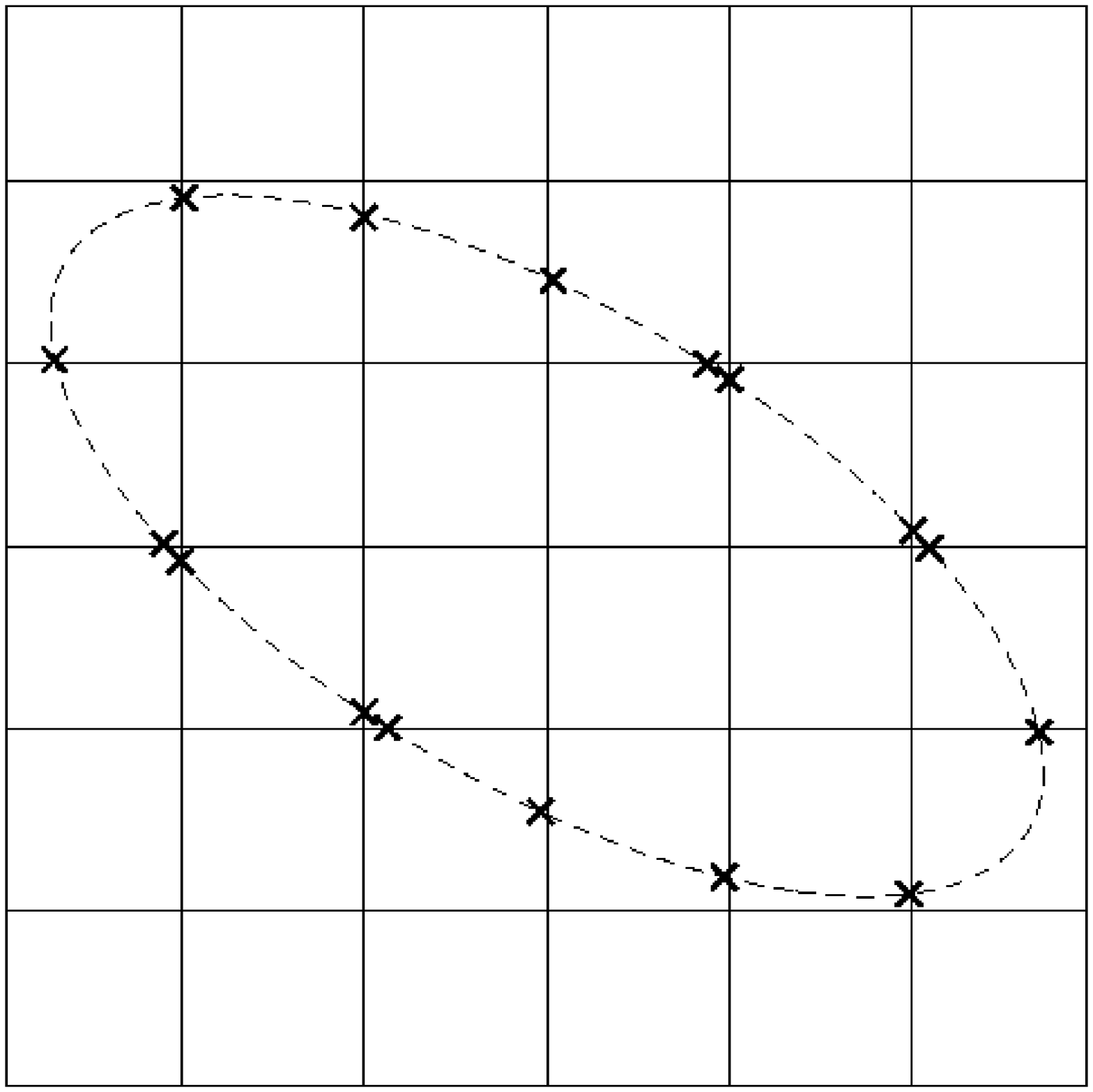}

\
      {\small {\bf{Fig.~1} \ }
      { Crossing points in the pixelized map for the smooth field ($h<<r_c$). Area
inside the ellipse corresponds to the region, where $f>\nu$.

}}}

\vspace{3cm}

The total number of these points in the map is:

\begin{equation}
N_{cr}=2\Delta N^2,
\end{equation}
where the 2 occurs in right hand side is because we use two directions for each pixel.

\begin{center}
{\it 3.2 Statistics of crossing points for signal + noise}
\end{center}

We consider uncorrelated Gaussian pixel noise independently
occurs  in  both components of polarization: $Q$ and $U$ with zero mean and
variance $\delta_o$. The resulting
signal in each pixel can be described as follows:

\begin{equation}
\begin{array}{l}
Q=Q_s+Q_n,\\
U=U_s+U_n,\\
\end{array}
\end{equation}
where indices s and n are for the signal and the noise respectively.
Therefore, one part of the signal is strongly correlated from pixel to pixel
(CMB) and another part (pixel noise) is completely uncorrelated. The Q and U components
of polarization obey the following relations:

\begin{equation}
\begin{array}{l}
\langle Q^2\rangle=\langle U^2\rangle=\sigma_o^2+\delta_o^2,\\
\langle (Q_2-Q_1)^2\rangle=\langle (U_2-U_1)^2\rangle=h^2\sigma_1^2+\delta_0^2,\\
h\sigma_1<<\sigma_0,
\end{array}
\end{equation}
where 1,2 denotes the values of $Q$ and $U$ in two neighbor pixels along one of the grid lines.
It is useful to introduce parameters: $a=\delta_0/\sqrt{\sigma_o^2+\delta_o^2}$ 
and $b=h\sigma_1/\sqrt{\sigma_o^2+\delta_o^2}$. We again use the dimensionless values:
$q=Q/\sqrt{\sigma_o^2+\delta_o^2}$, $p=P/\sqrt{\sigma_o^2+\delta_o^2}$. The 
probability, that the space between two neighbor pixels contains up-crossing or
down-crossing point is equal to the following integrals:

\begin{equation}
\begin{array}{l}
\Delta_q=2\int\limits_{-\infty}^qdq_1\int\limits_q^{+\infty}dq_2F_q(q_1,q_2),\\
\Delta_p=2\int\limits_{0}^pdp_1\int\limits_p^{+\infty}dp_2F_p(p_1,p_2)
\end{array}
\end{equation}  
for q and p values correspondingly. $F_q$ and $F_p$ are joint probability distribution
functions for $q$ and $p$ values in two neighbor pixels (1 and 2). Eq(10) has two
obvious asymptotics. If noise is much less than the signal ($a<<b<<1$), then
it is useful to make the substitution $\overline{q}=(q_1+q_2)/2$, $q_x=(q_2-q_1)/b$
and $\int d\overline{q}\overline{F_q}(\overline{q},q_x)=bq_x\overline{F_q}(q,q_x)$.
Analogous result is, of course, for the p field. 
Finally, we get the same formulae as in the previous subsection for the pure
signal.

On the other hand, if the noise is much bigger than the signal ($a\approx 1$), then
random values in neighbor pixels are independent: 

\begin{equation}
\begin{array}{l}
F_q(q_1,q_2)=\frac{1}{2\pi}e^{-\frac{q_1}{2}}e^{-\frac{q_2}{2}},\\
F_p(p_1,p_2)=p_1p_2e^{-\frac{p_1}{2}}e^{-\frac{p_2}{2}},
\end{array}
\end{equation}  
and we get a very simple result:
\begin{equation}
\begin{array}{l}
\Delta_q=2(1-\Phi(\frac{q}{\sqrt{2}}))\Phi(\frac{q}{\sqrt{2}}), \\
\Delta_p=2(1-e^{-\frac{p^2}{2}})e^{-\frac{p^2}{2}}.
\end{array}
\end{equation}  

The result of integration in (10) is quite complicated and can be found in the
appendix of the paper (Naselsky et. al. 2000). In (fig. 2) we demonstrate 
the number of crossing points for p
values as a function of the level for different values of a (b=0.07).

\parbox{4.5in}{
      {\includegraphics[scale=0.55,width=4in,totalheight=2.5in]{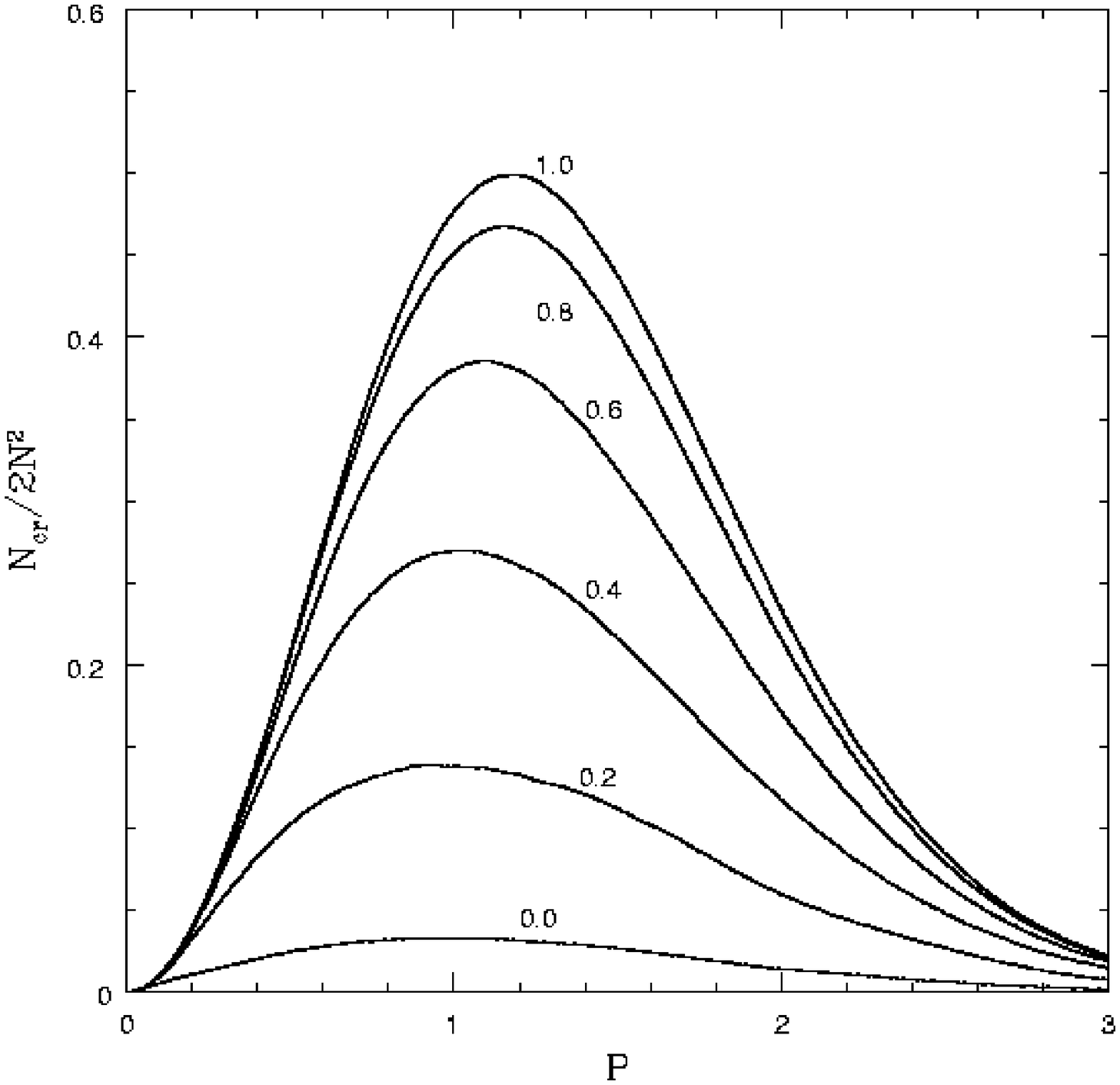} }

\
      {\small {\bf{Fig.~2} \ }
      { Number of crossing points in the pixelized map divided by the total
number of pixels as a function of level. Numbers indicate curves for different
values of a.

}}}

\vspace{0.5cm}

\parbox{4.5in}{
 \includegraphics[scale=0.55,width=4.5in,totalheight=2.5in]{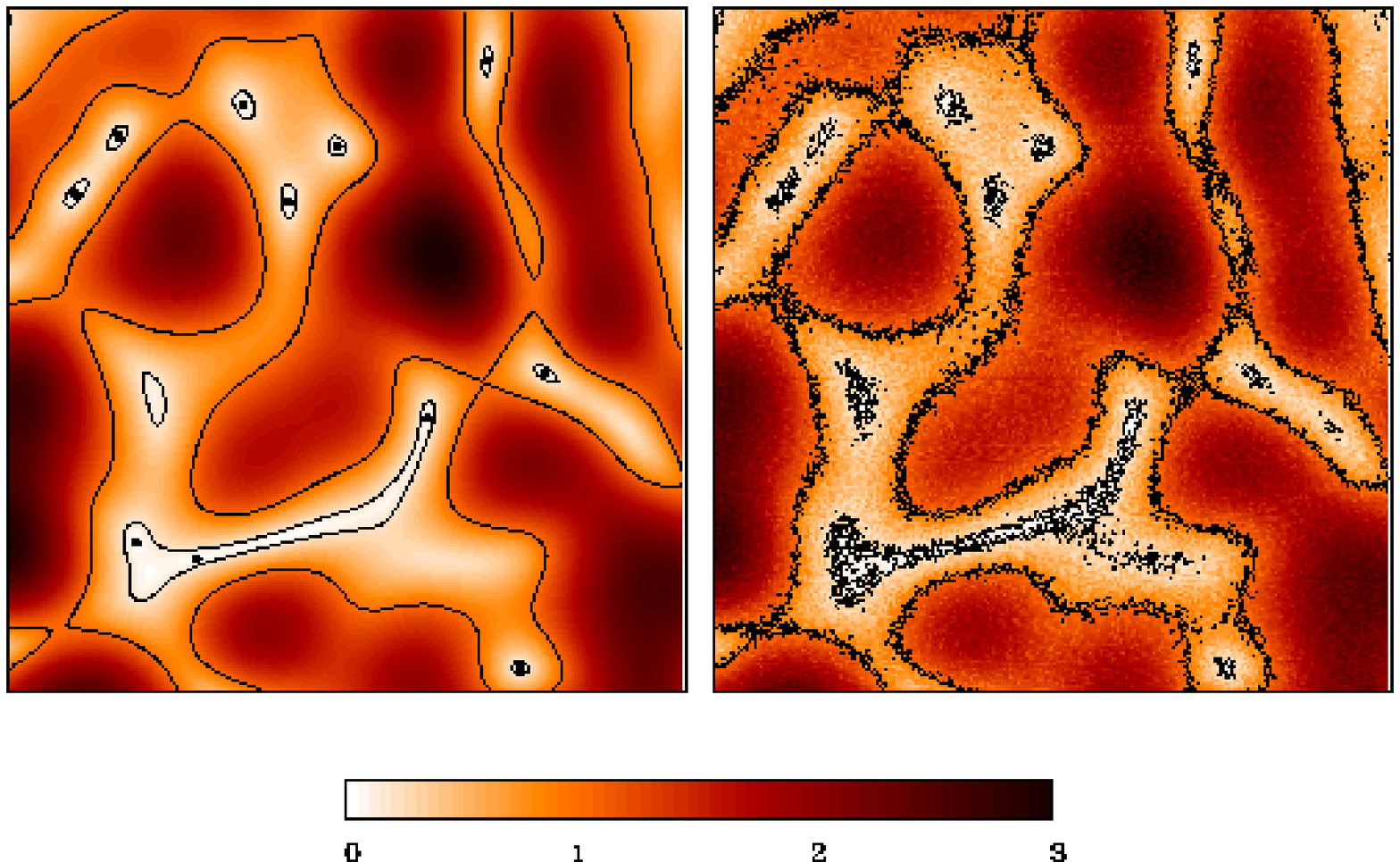}}

\
      {\small {\bf{Fig.~3} \ }
      { Left panel: pure CMB signal. Right panel: signal+noise. Small
points are crossing points at the levels p=0.2 and p=1. Shaded circles
show the positions of non-polarized points for the pure CMB signal.
}}

\vspace{1cm}

We have simulated $5^o\times5^o$, $256\times 256$ pixels 
map of CMB polarization for 
the standard CDM cosmological model and the same map with $10\%$ of the 
noise/signal ratio ($\delta_0/\sigma_0=0.1$) (fig. 3). 
In the second map one can see the 'non-zero width'
of the isopolatization lines. Number of crossing points is definitely higher
for the map with the noise. 

Finally, we suggest using the total number of crossing points
in the observational map for different levels in order to construct
the best fit line (Eq(10)) with parameters a and b. Therefore,
we can find parameters of the signal ($\sigma_0$, $\sigma_1$) and
noise ($\delta_0$) before the subsequent filtering. 

\begin{center}
{\it 3.3 Non-polarized points in the map}
\end{center}

Non-polarized (or singular) points in the map are the points where both components Q and U of
a pure CMB signal are equal to zero. These points of polarization are 
a natural part  of the geometrical structure for the CMB signal and 
their total number $N_{np}$ in the map is $\approx S/r_c^2$, where S is the total area of the map
(see for details (Naselsky and Novikov 1998)). Such points can be of three different types:
saddles, comets and beaks. Concentrations of singular points of different
types in case of a Gaussian signal are $0.5N_{np}$, $0.04N_{np}$ and $0.46N_{np}$
for saddles, beaks and comets correspondingly. Therefore, they can provide the
statistical information about the nature of the signal.

In addition to the mentioned properties, these points can be used for the analysis
of the noise in their vicinities. At the small distance r ($r<<r_c$) from such a point
signal is sufficiently small $P_s\approx r\sigma_1$. Therefore,  in the area, where
$r<\delta_0/\sigma_1$ the signal is much smaller, than the noise. Roughly speaking,
pixels inside this area indicate the noise only.
We suggest using this fact for the estimation of the pixel-pixel correlations in
the noise and (if they exist) making an appropriate correction in the final formula (10).  

\section{Conclusions}

In this paper we propose the method of determination of the spectral parameters
for the cosmological signal and pixel noise using the observational data of the
CMB polarization without any additional assumptions. We also suggest use of
singular points of the polarization for the same purpose.

To determine the parameters of the noise we introduced the notion of the 
crossing points in the observational maps (see section 2). We obtained the
formulae for the total number of them at different levels for 
the modulus of polarization P ($N_p$) and separately for two components Q and U
($N_{q,u}$) in case of presence of the CMB signal and pixel noise. This formulae 
includes parameters $\sigma_0$, $\sigma_1$, $\delta_0$. The theoretical 
expressions $N_p$, $N_{q,u}$ that fit the observational data allow us to determine
these parameters. On the other hand, the determination of correlations in the observational
data in the very vicinities of the points, where the CMB polarization
vanishes and where there is the noise only (spot like regions of crossing dots in
the map (fig.3)) allows us to correct the final result.

We would like to emphasize, that this approach is applicable for the 
analysis of the CMB anisotropy as well. 

This method allows us to estimate spectral parameters of signal and noise 
directly from the observational data without any additional assumptions before subsequent
filtration.

\begin{center}
Acknowledgments. 
\end{center}

This investigation was supported in part by the Danish Natural 
Science Research Council through grant No. 9701841 and also in part by the
grants INTAS-1192 and RFFI-17625 and by  
 Danmarks Grundforskningsfond through 
its support for establishment of the TAC. 
PN is grateful to Per Rex Christensen and H.U. Norgaard- Nielsen for
discussions. DN acknowledges
Board of the Glasstone Benefaction. 

\begin{center}
{\bf References} 
\end{center} 
.\\
Bouchet, F.R \& Gispert R., New Astron., 4, 443, (1999)\\
De Bernardis, P. et al. Nature, 404, 955 (2000)\\
Hanany, S. et al. astro-ph/0005123, (2000)\\
Naselsky, P. \& Novikov, D. 1998, ApJ, 507, 31\\
Naselsky, P., Novikov, D., Novikov, I. \& Silk, J., in preparation\\
Rees, M. ApJ, 153L1, (1968)\\
Schmalzing, J., Gorski, K.M. 1997, astro-ph/9712185\\
Tegmark, M. \& Efstathiou, G., MNRAS 281, 1297, (1996)

\end{document}